\documentclass[twocolumn]{aastex62}
\usepackage{amsmath}
\usepackage{comment}
\usepackage{xcolor}

\newcommand{\kepler}{{\it Kepler}}
\newcommand{\tess}{{\it TESS}}

\newcommand{\python}{{\tt PYTHON}}

\newcommand{\numpy}{{\tt NumPy}}
\newcommand{\RadVel}{{\tt RadVel}}
\newcommand{\forecaster}{{\tt forecaster}}
\newcommand{\BLENDER}{{\tt BLENDER}}
\newcommand{\multi}{{\tt MultiNest}}
\newcommand{\vespa}{{\tt vespa}}
\newcommand{\emcee}{{\tt emcee}}

\newcommand{\isochrones}{{\tt isochrones}}

\newcommand{\linalg}{{\tt linalg.lstsq}}
\newcommand{\RVLikelihood}{{\tt RVLikelihood}}
\newcommand{\logprob}{{\tt logprob}}
\newcommand{\likelihoodpy}{{\tt likelihood.py}}

\newcommand{\toitenb}{HD\,183579b}
\newcommand{\toiten}{HD\,183579}

\shorttitle{A WARM-NEPTUNE ORBITING A BRIGHT SOLAR ANALOG}
\shortauthors{Palatnick, Kipping \& Yahalomi}

\begin{document}


\title{VALIDATION OF HD\,183579b USING ARCHIVAL RADIAL VELOCITIES: \\
A WARM-NEPTUNE ORBITING A BRIGHT SOLAR ANALOG}

\correspondingauthor{Skyler Palatnick}
\email{skylerp@sas.upenn.edu}

\author[0000-0001-5053-2660]{Skyler Palatnick}
\affil{School of Engineering and Applied Sciences,
University of Pennsylvania,
Philadelphia, PA 19104, USA}

\author[0000-0002-4365-7366]{David Kipping}
\affil{Department of Astronomy,
Columbia University,
550 W 120th Street,
New York, NY 10027, USA}

\author[0000-0003-4755-584X]{Daniel Yahalomi}
\affil{Department of Astronomy,
Columbia University,
550 W 120th Street,
New York, NY 10027, USA}



\begin{abstract}
As exoplanetary science matures into its third decade, we are increasingly
offered the possibility of pre-existing, archival observations for newly
detected candidates. This is particularly poignant for the \tess\ mission,
whose survey spans bright, nearby dwarf stars in both hemispheres - precisely
the types of sources targeted by previous radial velocity (RV) surveys. On this
basis, we investigated whether any of the \tess\ Objects of Interest (TOIs)
coincided with such observations, from which we find 18 single-planet
candidate systems. Of these, one exhibits an RV signature that has the
correct period and phase matching the transiting planetary candidates with a
false-alarm probability of less than 1\%. After further checks, we exploit this
fact to validate \toitenb\ (TOI-1055b). This planet is $<4$\,$R_{\oplus}$ and has better than 33\%
planetary mass measurements, thus advancing \tess’ primary objective of
finding 50 such worlds. We find that this planet is amongst the most
accessible small transiting planets for atmospheric characterization. Our work
highlights that the efforts to confirm and even precisely
measure the masses of new transiting planet candidates need not always depend on
acquiring new observations - that in some instances these tasks can be
completed with existing data.
\end{abstract}


\section{Introduction}
\label{sec:intro}

Large scale photometric surveys have revolutionized our understanding of
extrasolar planets through the discovery of thousands of transiting planets
\citep{batalha:2014}. Despite the many advantages of the transit method, it
suffers from one primary weakness - a planet-like transit can be caused by
numerous false-positives \citep{bryson:2013,santerne:2013,leuquire:2018}.

Determining the true nature of a planet-like transit is complicated by the
fact that the false-positive rate (FPR) varies between different photometric
surveys; for example, for \kepler\ it is ${\sim}10\%$ \citep{borucki:2011}
but is far higher for ground-based surveys like KELT \citep{collins:2018}.
Of particular interest for this work is that \tess\ is predicted to have
a FPR of ${\sim}40$\% for the 2-minute cadence targets \citep{sullivan:2015}.
However, even when one focusses on a specific survey, the FPR varies
dramatically with the planetary radius; for example, \kepler's FPR is 17.7\%
for giant planets but 6.7\% for Neptunes (\citealt{fressin:2013}, but also see
\citealt{santerne:2012}). Yet more, it appear to further depend on the
evolutionary state of the parent star (\citealt{sliski:2014}, but also see
\citealt{gaidos:2013}) and even position within the detector's pixel array
\citep{christiansen:2020}.

The ``gold standard'' solution for distinguishing between genuine transiting
planets and false-positives is to obtain radial velocity (RV) measurements
that detect the presence of a Doppler signal of the same period and phase as
the transit signal (e.g. \citealt{hebrard:2019}). In the era of thousands of
planetary candidates dawned by \kepler, this approach remains powerful
but ultimately limited in impact due to the challenge of securing the very
large number of observing nights required to cover the entire sample. High
significance RV detections are typically described as planet ``confirmations''
in the associated paper (e.g. \citealt{almenara:2018}). In some cases, the
lack of a detectable RV signal in phase with a known transiter has been used to
place mass upper limits, which is then used as a basis to describe said
transiter as ``confirmed'' (e.g. \citealt{timmermann:2020}.

Due to the imbalance of RV observations versus transiting planet candidates,
alternative strategies have been developed in recent years to make further
progress. Specifically, the community has become familiar with the concept of
``statistical validation'' of planetary candidates. Unlike confirmations,
which are implied to be essentially certain planets, validations frame
signals as being planets to some probability threshold. These validations
generally consider information such as the transit morphology, centroid
positions and high resolution imaging constraints in the context of both planet
and non-planet scenarios (e.g. hierarchical triples), in order to quantify the
odds of planethood.

Such validation efforts find their early footing in the work of
\citet{torres:2004}, who showed that the transiting planet candidate
associated with OGLE-TR-33 was likely a false-positive. This technique was
applied the following year to OGLE-TR-56, which finally led to the first
statistically validated planet \citep{torres:2005}.

Statistical validation of transiting planet candidates remained somewhat
of a niche exercise in the exoplanet community during these years. This was
largely because most transiting planet candidates were coming from ground-based
surveys, such as HATNet \citep{bakos:2004} and WASP \citep{pollacco:2006},
whose targets were bright and not so numerous that RV confirmation was almost
ubiquitous in the resulting papers. This situation shifted in the \kepler-era,
when exoplanet astronomers began drinking from the fire hose and could no
longer keep up with the large catalogs of planetary candidates being released
(e.g. \citealt{borucki:2011}). The \BLENDER\ software \citep{torres:2011}
was the first attempt to generalize the validation framework for en-masse
work and led to the validation of dozens of new \kepler\ planets
\citep{torres:2015,torres:2017}. In tandem, more computationally
efficient validation methods were developed, such as \vespa\
\citep{morton:2012} and Gaussian Process Classification
\citep{armstrong:2020}, enabling the validation of many hundreds of \kepler\ 
candidates. The inclusion of planet multiplicity
information into the validation methodology by \citet{lissauer:2012} further
empowered the approach to the point where the majority of \kepler\ planetary
candidates have now been validated \citep{rowe:2014,morton:2016}.

Despite the apparent dwindling need for RVs during these years, the value of
RV measurements is arguably entering somewhat of a renaissance. Precise
RV data sets have now been accumulating for two decades \citep{butler:2017}
revealing the long-period population of exoplanets (e.g.
\citealt{wittenmyer:2016}). Yet more, the paucity of observed planet masses
amongst the transiting population has highlighted their need, such as
better constraints on the planetary mass-radius relation \citep{chen:2017},
predicting the scale-height in transmission spectroscopy work
\citep{anglada:2013}, and removing degeneracies in exomoon work
\citep{teachey:2018}. Yet more, the impact of RVs is buoyed by the fact
that \tess\ is focussed on brighter stars \citep{ricker:2015}, which are far
more amenable to RV characterization than those of \kepler. Indeed, it is
telling that a primary goal of the \tess\ mission is to measure the masses
of 50 small ($\lesssim4$\,$R_{\oplus}$) exoplanets (despite the fact
\tess\ itself generally cannot measure masses).

In the \tess\ era, then, RVs will clearly play a more impactful role than that
of \kepler, and, in this work, we consider to what extent they can be used to
validate known transiting planet candidates. In particular, we turn to publicly
available archival RV surveys over the last two decades of the northern and
southern skies which include many targets not just observed by \tess\ but found
to have TOIs (\tess\ Objects of Interest). Although no planets may have been
detectable in the original time series, the inclusion of the \tess\ ephemerides
(which are typically very precisely measured) adds new constraints to these
data sets that may elevate signals previously lying beneath the noise floor. 
We note that \citet{huang:2018} utilized similar methodology to validate 
HD\,39091c, but their approach differs in that it does not generalize to be 
applicable to the entire TOI database, and in our work we push into lower signal-to-noise ratios.

We describe our new approach in Section~\ref{sec:observations}, along with
the identification of one newly validated planet. In
Section~\ref{sec:transits}, we explore the physical properties of this planet
by including the constraints from the transit light curve and stellar
isochrones. The importance of this individual planet is discussed in 
Section~\ref{sec:discussion}, along with a broader brush discussion of this
new approach to validating exoplanets.

\section{Radial Velocity Analysis}
\label{sec:observations}

\subsection{Cross Referencing TOIs}
\label{sub:crossref}

We begin by curating a list of sources which have RV measurements available,
looking in particular for sources which have had no
previous planet detections. Although numerous surveys have been published over the
years, we limit ourselves to just the largest surveys in order to provide a
degree of catalog homogeneity. Specifically, we seek one large survey in each
celestial hemisphere to provide the necessary data. To this end, we identify
the Lick-Carnegie Exoplanet Survey (LCES) using the HIRES instrument on Keck-I,
and the High Accuracy Radial Velocity Planet Searcher (HARPS) mounted on the
3.6m ESO telescope at La Silla as most suitable.

From LCES, we obtained 60,949 publicly available radial velocities for 1624
unique sources processed and published by \citet{butler:2017}. These
observations span 20 years with an instrument upgrade occurring in August 2004.
Despite this, \citet{butler:2017} report no significant velocity offset after
the upgrade in their published RVs, and explicitly state so in their work, and thus we will treat the entire data set
as originating from a single instrument.

From HARPS, we obtained over 212,000 RVs for 2912 sources
\citep{trifonov:2020}. HARPS has been mounted since 2003 but an instrument
upgrade in May 2015 introduces a RV offset that needs to be accounted for
between these two eras, and one which is different for each star
\citep{trifonov:2020}.

In rare cases, sources were caught by both surveys. Of the 6 unique sources for
which this was the case, 4 were already known planet detections, and thus were
not impactful to our overall procedure. The remaining 2 were treated in the
same manner as the HARPS upgrade, with an RV offset that needs to be accounted
for between the two instruments which is different for each star.

We then proceeded to filter the list down to only sources which were also
listed as a TOI via the TESS Alert system, yielding 100 TOIs from 97 sources.
Of these, 70 were already known planet detections at the time of writing and so
these were excluded from our analysis. We also excluded 6 TOIs that had 5, or
fewer, total RV observations, and 3 sources with two TOIs each in the same
system because multi-planet systems are not compatible with our validation
methods. For multi-planet TOIs, our validation methodology can not adequately disentangle the signal between the two planet candidates, and thus their inclusion in our analysis is unnecessary. This provides us with a final list of 18 TOIs which have not been confirmed as planets as of the time of writing, and have archival precise
radial velocities available from either HIRES or HARPS. These 18 TOIs are
listed in Table~\ref{tab:FAPtable}.

\begin{table*}
\caption{
Summary of several tests applied to our TOIs to identify statistically significant
and physically sound radial velocity solutions.
$\dagger$ = An outlier point was removed during the analysis.
} 
\begin{center}%
\begin{tabular*}{\textwidth}{@{\extracolsep{\fill}}llllllll}\hline
TOI & Main Identifier & FAP &
$K_{\mathrm{\RadVel}}$\,[m/s] & $K_{\mathrm{\forecaster}}$ [m/s] & $K_{\mathrm{circ}}$ [m/s] &
LS test for $K_{\mathrm{\RadVel}}$ & Physicality $p$-value
\\\hline
1055.01 & \toiten & $0.32\%$ & $4.7_{-1.2}^{+1.1}$ & [1.2, 8.1] & 3.8 & 0.49\% & 0.23 \\
260.01 & GJ\,1008 $\dagger$ & $1.2\%$ & $3.44_{-0.80}^{+0.78}$ & [0.0, 5.1] & 3.1 & $\cdots$ & $\cdots$ \\
560.01 & GJ\,313 & $1.48\%$ & $14.2_{-2.5}^{+3.0}$ & [2.0,12.3] &19 & $\cdots$ & $\cdots$ \\
1611.01 & HD\,207897 & $2.6\%$ & $7.3_{-3.5}^{+4.7}$ & [0.6, 4.3] & 2.8 & $\cdots$ & $\cdots$ \\
1827.01 & Wolf\,437 $\dagger$ & $4.2\%$ & $4.6_{-2.8}^{+2.3}$ & [0.9, 9.7] & 2.5 & $\cdots$ & $\cdots$ \\
1011.01 & HD\,61051 & $9.1\%$ & $3.04_{-0.74}^{+0.79}$ & [0.6, 3.5] & -1.2 & $\cdots$ & $\cdots$ \\
179.01 & HD\,18599 & $9.4\%$ & $36.4_{-7.4}^{+7.5}$ & [1.5, 8.5] & 18 & $\cdots$ & $\cdots$ \\
440.01 & HD\,36152 & $13.0\%$ & $1.7_{-1.7}^{+1.2}$ & [1.5, 6.8] & -0.55 & $\cdots$ & $\cdots$ \\
461.01 & HD\,15906 & $21.5\%$ & $-9.3_{-7.8}^{+7.7}$ & [1.3, 5.2] & 2.6 & $\cdots$ & $\cdots$ \\
1860.01 & HD\,134319 & $34.7\%$ & $132_{-62}^{+50}$ & [0.0, 10.3] & 59 & $\cdots$ & $\cdots$ \\
486.01 & GJ\,238 & $50.6\%$ & $0.57_{-0.73}^{+0.72}$ & [0.06,0.18] & 0.61 & $\cdots$ & $\cdots$\\
909.01 & HD\,150139 & $57.8\%$ & $2.3_{-1.5}^{+1.0}$ & [0.8, 8.3] & -0.52 & $\cdots$ & $\cdots$ \\
198.01 & GJ\,7 & $63.6\%$ & $39.5_{-12.0}^{+4.3}$ & [0.5, 3.4] & 27.8 & $\cdots$ & $\cdots$\\
1970.01 & TYC\,8647-2057-1 & $78.7\%$ & $6600_{-201}^{+220}$ & [42, 18000] & -71 & $\cdots$ & $\cdots$\\ 
253.01 & HIP\,4468 & $87.1\%$ & $-32_{-32}^{+39}$ & [0.0, 8.0] & 0.77 & $\cdots$ & $\cdots$\\
731.01 & GJ\,367 & $89.5\%$ & $0.28_{-0.84}^{+0.63}$ & [0.0, 4.6] & 0.11 & $\cdots$ & $\cdots$\\
139.01 & HIP\,110692 & $91.2\%$ & $5.2_{-2.6}^{+2.5}$ & [1.4, 6.7] & 0.45 & $\cdots$ & $\cdots$ \\
741.01 & GJ\,341 & $92.6\%$ & $0.22_{-0.49}^{+0.45}$ & [0.0,2.0] & 0.054 & $\cdots$ & $\cdots$\\ [1ex]
\hline
\label{tab:FAPtable}
\end{tabular*}
\end{center}
\end{table*}

\subsection{A Check for Long-term Trends}
\label{sub:trends}

Before we can look for the short-period RV signals expected due to the TOIs, it
is necessary to check for evidence of long-term trends in the data. If these
should exist, a failure to account for them would degrade our sensitivity to
detect low amplitude signals. To accomplish this, we performed a linear
least squares fit of the RV time series using the inverse square of the
reported uncertainties as the weights (no jitter is included for this test). A
flat, linear and quadratic trend model are regressed to each time series, from
which we compute a $\chi^2$ and BIC \citep{schwarz:1978} score. The model with
the lowest BIC is saved as the appropriate trend model for each TOI. To account
for the effect of the 2015 upgrade to HARPs on RV datasets with observations
that span the eras before and after 2015, we implemented a Nelder-Mead
minimization routine to solve piecewise equations accounting for the
offset between the RVs from the pre- and post-upgrade time periods
corresponding to flat, linear, and quadratic trends. We then follow the stated
procedure of choosing the model with the lowest BIC as the correct trend for
the given TOI. The same workflow was applied to the TOIs with data from both
HARPS and LCES. We note that while the trend models we adopt are favored by the BIC score for a given TOI, these trends are not necessarily statistically significant. Of the TOIs we determined to have RV data with a linear or quadratic favored trend model, TOIs 486.01, 560.01, 741.01, 198.01, 1860.01, 909.01, 1055.01, 179.01, 1611.01, 440.01, 1011.01, 253.01, and 1970.01 had $\Delta\mathrm{BIC}>10$, indicating a strong likelihood of a trend in their respective RV data.

\subsection{Calculating False-Alarm Probabilities (FAPs)}
\label{sub:FAPs}

We considered several tests to evaluate whether there is a genuine RV signal in
the archival data associated with each TOI, but the first of these is a
bootstrapping false-alarm probability (FAP) test. This test consists of three
steps, described as follows. First, for each TOI, we fit a trend + circular
orbit model (which can be expressed as a purely linear model for a given
period) to the TOI's RV data, weighting by the inverse square uncertainties,
and evaluate the $\chi^2$ goodness-of-fit. The trend component of the model corresponds to the BIC best fit trend; if the BIC for a given RV data set favors a flat trend, only a constant offset term is included, while if the BIC for a given RV data set favors a linear or quadratic trend, linear or quadratic terms are included in addition to the constant offset term. We allow for negative $K$ values
during this process, which can be used a diagnostic for ``bogus'' detections
later. For planets on near circular orbits, which is broadly expected given the
short-period nature of the TOIs, one expects the phase folded RVs to follow an
inverted sinusoid of amplitude $K_{\mathrm{circ}}$ \citep{beta:2013}. What this means is that at
the time of inferior conjunction (i.e. mid-transit time), the RV signal should
be zero since the star is moving tangentially, but the acceleration should be
blueshifting maximally (i.e. the RV gradient is maximally negative). Thus, our
linear equation was of the form:

\begin{align}
\mathrm{RV}(t) = a_{0} + a_{1}(t-t_{0}) + a_{2}(t-t_{0})^2
                 - K_{\mathrm{circ}} \sin(\frac{2\pi{(t-\tau)}}{P})
\label{eq:linearfit}
\end{align}

\noindent
where $t_{0}$ is a pivot point selected near the mid-point of the observational
baseline, and $a_{2}$, $a_{1}$, and $a_{0}$ are constants corresponding to
quadratic, linear, and constant trends in the data, respectively. We utilized
the \linalg\ function from the \numpy\ \python\ module to regress
Equation~(\ref{eq:linearfit}) to the available RVs for each TOI. Once again,
for TOIs with HARPS RV data that spans the pre and post upgrade eras as well as
the TOIs with data from both HARPS and LCES, we used the non linear Nelder-Mead
minimization routine to solve the piecewise equation accounting for the
constant offset between the two observation periods or the two different
instruments.

Since this is a fit, with some parameter flexibility, then the resulting
$\chi^2$ will always be better than that obtained without the sinusoid present.
Thus, an improvement in $\chi^2$ is not sufficient to claim a detection.
Further, the noise properties cannot be assumed to behave as strictly Gaussian
and thus we avoid making detection claims based on the degree to which $\chi^2$
improves either. In light of these points, how can one go about evaluating a
probability for the reality of these signals?

We approach this through bootstrapping. Specifically, we repeat the same
procedure described above but with a different (and ultimately false)
ephemeris. The orbital period is drawn from a probability distribution which
approximates the observed TOI period distribution, but we exclude any periods
which are within 20$\%$ of the true answer. This approximate distribution
was found by first inspecting the distribution the log-periods of the 2330
available TOIs, which exhibit an approximately triangular distribution mixed with
a background uniform distribution. We performed likelihood maximization of a
uniform+triangular mixture model, with support defined over the range of
the available log-periods, yielding a mixture model which is 0.777 triangular,
whose shape has a mode at $\log P = 0.34$, a minimum at $-0.026$ and a
maximum at $3.804$. After a period is selected from this distribution, the
phase is simply randomized uniformly. For each random ephemeris, a linear
equation with an inverted sinusoid and trend is fit (or non-linear piecewise
equation for the TOIs where this is necessary), and the $\chi^2$ improvement
is recorded. Since the real fit allows negative $K$ values, the exact same
procedure and rule set is used for the bootstrap to keep everything
like-for-like.

A false-alarm probability (FAP) score can then be computed by asking, how often
do the fake ephemerides lead to a $\chi^2$ improvement which is superior than
the improvement obtained with the true ephemeris? This can be quantified using
a one-tailed $p$-value then, similar to the typical bootstrapping applied to
periodogram analyses in RV surveys. RV signals driven by stellar activity can occur across a broad range of frequency space and, in general, have no reason to coincide with a series of periodic and statistically significant box-shaped dips that represent a TOI. In this way then, by evaluating the power at different random but representative ephemerides, our FAP scores inflate in the presence of such behavior. A consequence of this is that our approach may obtain false-negatives, genuine RV planets that we reject because there is an activity signal present. Nevertheless, we prefer to err on the side of being conservative in this sense when validating planets in this work.

\begin{figure}
\includegraphics[width=\columnwidth]{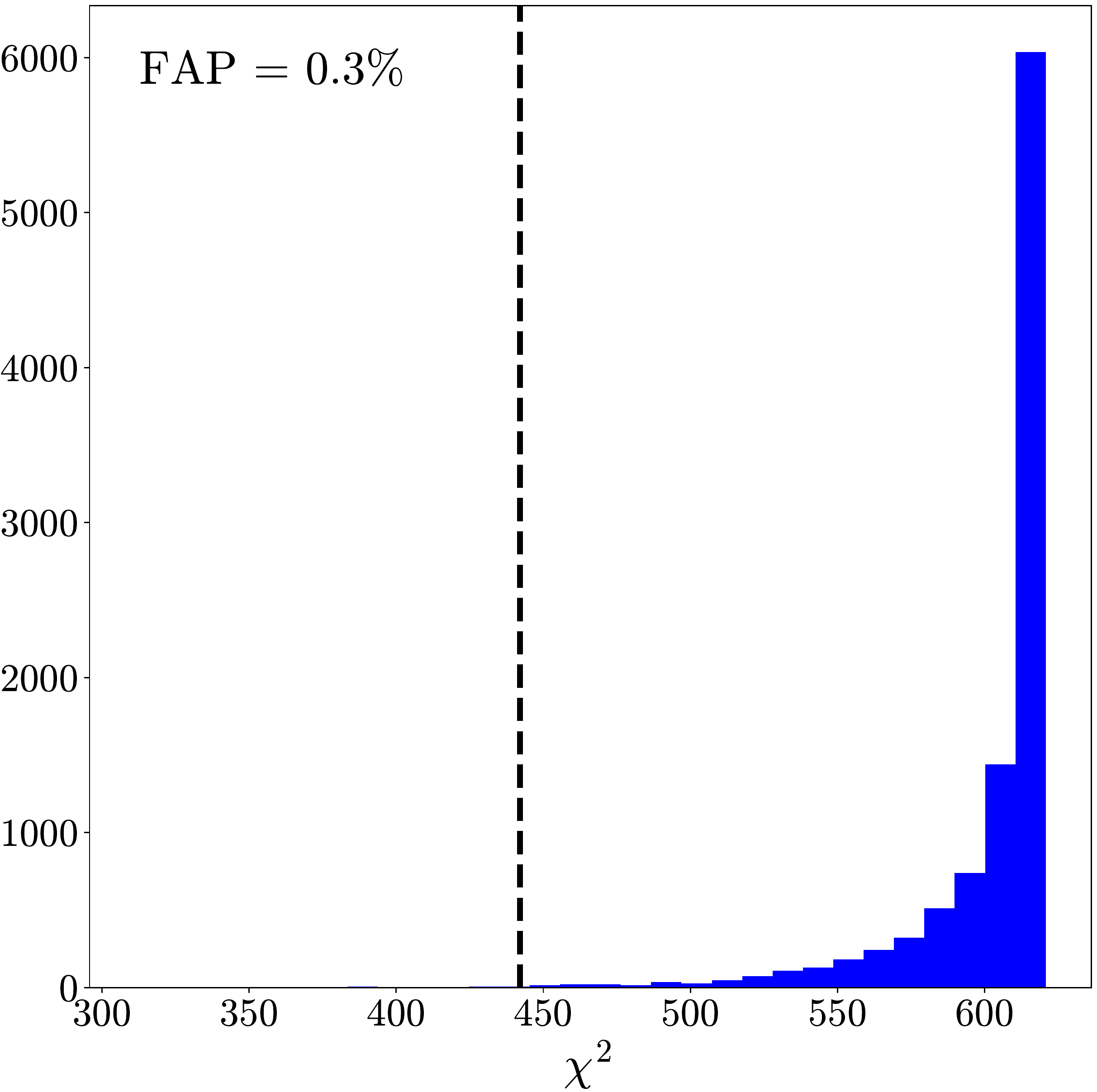}
\caption{Results of the FAP test for \toiten.01. The FAP percentage is reported in the upper left corner, and the $\chi^2$ value for each true linear fit (using the real $P$ and $\tau$) is denoted by a dashed black line.}
\label{fig:fakePplots}
\end{figure}

Following \citet{morton:2016}'s validation
work on \kepler\ transiting planets, we consider any FAP score lower than 1\%
grounds for potential validation (subject to some further checks and tests).
The FAP scores are listed in Table~\ref{tab:FAPtable}, where 1 of the 18 TOIs
exhibits a FAP below 1\% - TOI-1055.01 (HD\,183579.01).
Figure~\ref{fig:fakePplots} includes a histogram with the results of the FAP 
test for this planet.

Four other TOIs exhibit FAP scores below 5\%. In this work, these TOIs will
not be considered further as candidates for validation, but we note that they
are likely planets. These are TOI-260.01 (GJ\,1008.01), TOI-560.01
(GJ\,313.01), TOI-1611.01 (HD\,207897.01), and TOI-1827.01 (Wolf\,437.01).

\subsection{\RadVel\ Modeling and Testing for Non-Zero Semi-Amplitudes}
\label{sub:radvelfits}

To confirm the validity of this TOI as a planet, we conducted a more
thorough analysis of its RV data and the light curve of its
host star and then ran two additional tests. In total, 54 RV measurements of
\toiten\ were taken by HARPS over the course of 5.5 years.
To analyze these RVs, we utilize the \RadVel\ package \citep{fulton:2018}.

\RadVel\ uses MCMC regression to fit for various physical parameters including
$P$, $\tau$, $e$ (eccentricity), $\omega$ (argument of periastron), $K$, and RV
jitter. Since the object is transiting, it will be subject to the eccentricity
bias affecting transiting bodies due to geometric probability
\citep{barnes:2007,burke:2008}. This can be formally accounted for by using the
$e$-$\omega$ joint prior of \citet{eprior}, specifically their Equation~(23).
However, the prior is unstable for an intrinsically uniform prior in $e$,
described in $\alpha=1$ and $\beta=1$ in that expression. Instead then,
we use a Beta distribution intrinsic prior for $e$ of $\alpha=1$ and
$\beta=2$. Formally, the RV population of short period planets is better
described by $\alpha=1$ and $\beta=3$, which places more emphasis on
low-eccentricity orbits. We elect to $\beta=2$ to create a softer, flatter
and more uninformative prior, yet one which is stable and gently favors
more circular orbits. The intrinsic prior on $\omega$ is uniform over the
the $2\pi$ interval.

Since $P$ and $\tau$ are strongly constrained from the transit light curve, we
employ a Gaussian prior on these terms at their respective \tess\ ephemeris
values. For the RV jitter parameter, $\sigma_{\mathrm{jitter}}$, we employ a
broad log-uniform prior from 10$\%$ of the median RV error up to twice the
range of the RV data. For $\gamma$ (RV offset), $\dot{\gamma}$ (RV drift), and $\ddot{\gamma}$ (RV curvature), we employ the default \RadVel\ settings of a uniform prior with initial guesses of the median RV value, 0m\,s$^{-1}$, and 0m\,s$^{-1}$, respectively. The bounds on these terms are set by the range of
the RV data in hand. For $K$, we wanted to ensure zero-valued and negative solutions were
free to be explored and so we adopt a uniform prior from zero minus twice the range of the RV data to zero plus twice the range of the RV data. The upper and lower limits on this prior are chosen to simply allow any detectable signal with a period shorter than the baseline to be modeled by \RadVel.

Our \RadVel\ fits use the default mode of running 8 independent ensembles in
parallel with 50 walkers per ensemble for up to a maximum of 10000 steps per
walker, or until convergence is reached (See \citealt{fulton:2018}
for further details.) We also inspected the posteriors to
check for convergence and mixing and then use them in our calculations of
physical properties, along with the transit posteriors. We make these
posteriors publicly available at
\href{https://github.com/skypalat/toi_validation}{this URL}.

Once the fits were finished, we conducted two basic checks on the marginalized
posterior distribution for $K$. First, if the median of the distribution is
negative, the TOI was discarded, which occurred for TOI-461.01 and TOI-253.01.
However, we note that neither of these had low FAPs and thus would
have been rejected regardless. Second, for our TOIs with a FAP score below
1\% (which is just TOI-1055.01), we wanted to test if the $K$ posterior was
significantly pulled away from zero, implying a positive detection. Using
Bayesian evidences is somewhat unsatisfactory here because those values would
strongly depend upon the width of our prior. In particular, for $K$, there is
no obvious upper limit and thus it can be increased arbitrarily and thus dilute
the Bayesian evidences. Instead, we argue that a better test is the classic
Lucy-Sweeney test to evaluate if a parameter is offset from zero
\citep{lucy:1971}. The Lucy-Sweeney test returns a FAP that the parameter in
question is consistent with zero, which we report in the penultimate column of
Table~\ref{tab:FAPtable}. The 1\% FAP planet validation threshold of
\citet{morton:2016} is again used as a minimum threshold (in addition to the
previous tests) for a candidate to be considered validated.

\subsection{Statistical Validation of \toitenb}

\begin{figure}
\includegraphics[width=\columnwidth]{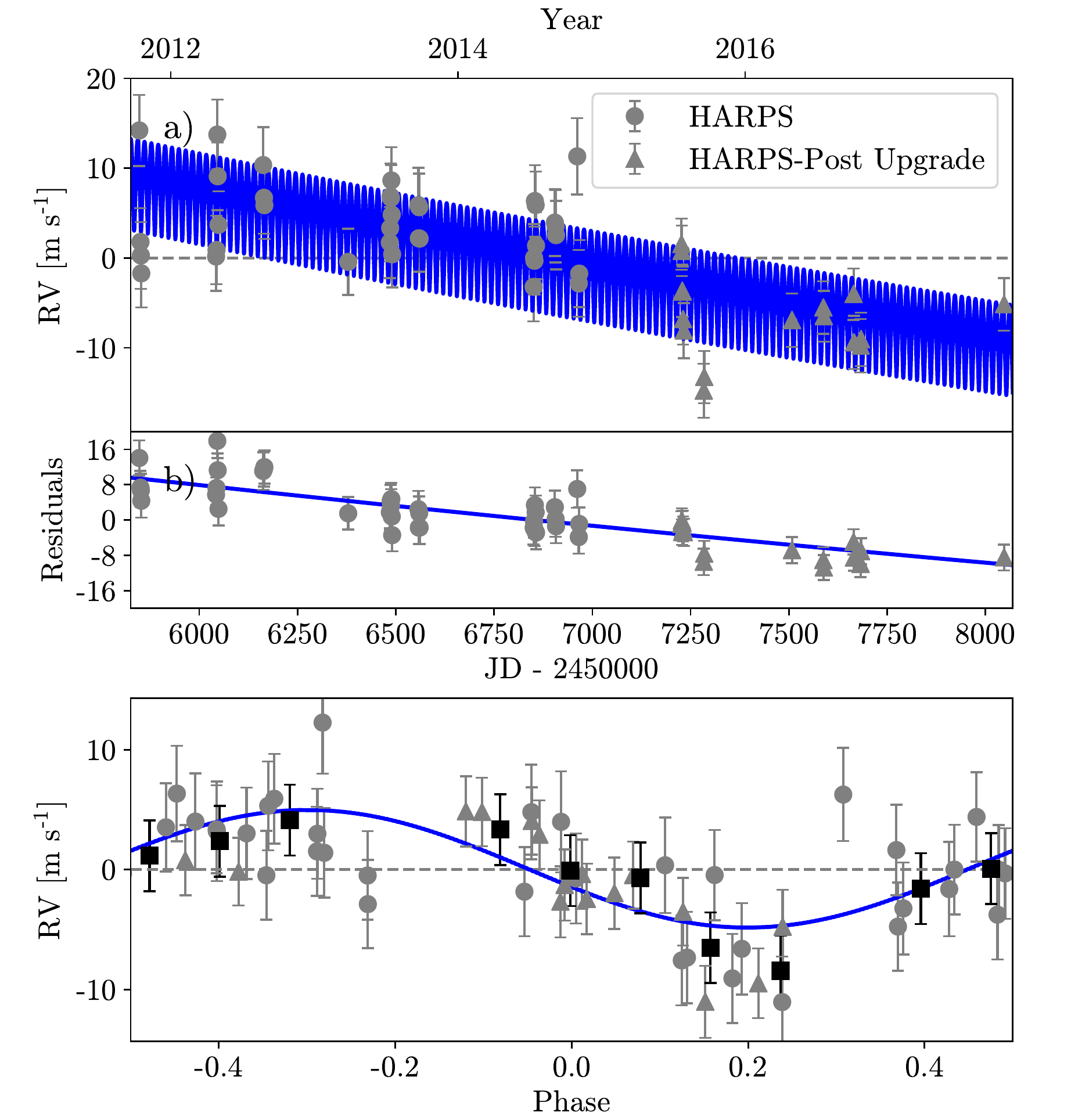}
\caption{
The RV data fit by \RadVel\ for \toitenb\ (TOI-1055b). The first and second
subplots from the top show \RadVel's fit to the raw data and the residuals
of those fits. The bottom plot from the top show \RadVel's fits of the phase
folded RV data. The black points in the phase fold plots represent binned data.
}
\label{fig:radvel1055}
\end{figure}

The TOI found earlier (see Section~\ref{sub:FAPs}) to exhibit
$<1$\% FAPs with the Monte Carlo test also exhibits a
$<1$\% FAP with the Lucy-Sweeney test (see Table~\ref{tab:FAPtable}), as
well as positive median $K$ values. Thus the \RadVel\ solution indicates
positive statistical evidence for a signal at the \tess\ ephemerides
for TOI-1055.01. The corresponding RV curves are shown in
Figure \ref{fig:radvel1055}.

As an additional check, we evaluated the one-sided $p$-value of the
\textit{a-posteriori} median $K$ value against the predictions from
\forecaster\ - just to ensure the fits are physically plausible. Since
\forecaster\ is an empirical mass-radius relation, then this essentially
asks whether the implied planetary densities are in the range one would
expect for a planet of its size. Once again, this TOI did not have a
suspicious $p$-value (less than 0.05) and thus appears physically sound.

We also re-visited the FAP calculation with consideration of the trend model
used. The existence of either a linear or quadratic trend appears
statistically secure with $\mathrm{BIC}_{\mathrm{quad}}=636.7$ and
$\mathrm{BIC}_{\mathrm{lin}}=641.7$, but $\mathrm{BIC}_{\mathrm{flat}}=672.8$, indicating that $\Delta\mathrm{BIC}>30$.
Although the quadratic model is favored over the linear model, we repeated
the FAP calculation with a linear model only, and obtained an even better
FAP score of 0.22\%. The low FAP score thus appears robust between these
two competitive trend models.

Finally, we examined the data validation (DV) reports for this TOI in order to
confirm that there were no indicators of false positive signals. DV reports are
generated by the NASA Science Processing Operations Center (SPOC) pipeline
\citep{2016SPIE.9913E..3EJ} for threshold crossing events (TCEs) in short
cadence observations and by the MIT Quick Look Pipeline (QLP,
\citealt{huang:2020}) for TCEs in long cadence observations (full frame
images). TOI-1055.01 has public DV reports from both the SPOC pipeline and the
QLP pipeline. We checked these DV reports to look for red flags such as
centroid offsets, differences in odd and even transits, or correlation between
the flux depth and the aperture size. We again find no reason to suspect the
transit signals to be spurious. We also note that the transit signal was
independently inspected by \citet{giacalone:2020} who find a 2\% FAP from
the light curve, not enough to validate but again indicating a likely real
planet.

From the passing of these checks in combination with our FAP scoring, we
conclude that TOI-1055.01 (HD\,183579.01) and is most likely a real planet to
$>99$\% confidence and refer to it as statistically validated in what follows
- thus updating its moniker to \toitenb. 

\subsection{A Note on Outliers}

We note that for several TOIs, the omission of outlier RVs has a noticeable
impact on their corresponding FAP score and Lucy-Sweeney results. RVs
considered as outliers are at least $6\sigma$ from both the favored long-term
trend model and the trend + circular fit, and also have large error bars
compared to the other RVs. TOI-253.01, while quite far from being validated,
saw its FAP score improve by 20\% with the omission of an outlier RV.
TOI-260.01 and TOI-1827.01 were just on the threshold of validation with the
exclusion of one outlier RV each, but not quite past the $>99\%$ benchmark.
Nonetheless, these two objects remain highly interesting and with further
observation, may prove to be planets.  

\section{Transit and Isochrones Analysis}
\label{sec:transits}

\subsection{Stellar Isochrones}
\label{sub:isochrones}

To complete our picture of the \toiten\ system, we require fundamental stellar
parameters. To this end, we performed a stellar isochrone analysis using the
\isochrones\ package by \citet{morton:2015}. The \isochrones\ package takes
the observable stellar properties as inputs and uses these to derive
fundamental properties by matching to stellar evolution models - in our
case we employed the Dartmouth models \citep{dotter:2008}.

As inputs, we start with the apparent magnitude in $V$-band reported in
\citep{koen:2010} and \citep{hog:2000}, and in the 2MASS $J$, $H$, $K$
bands by \citet{cutri:2003}. Next, we searched the literature for stellar
atmosphere parameters and elected to use the precise atmosphere parameters
reported in \citet{luck:2018}, which leverage the public HARPS spectra.

We also used the Gaia DR2 parallax from \citet{luri:2018} as a luminosity
indicator. This was included as an extra constraint on the stellar luminosity in the isochrone fits \citep{bakos:2010}.
Although our target is bright by exoplanet standards, the
brightest in $G$ for \toiten\ is 8.5 and is thus significantly fainter than
the $G\lesssim5$ range highlighted by \citet{drimmel:2019} as exhibiting
strong biases. However, we do account for the much smaller systematic parallax
error reported by \citet{stassun:2018}. All of these input parameters are
listed in Table~\ref{tab:table_stellar}. We ran \isochrones\
\citep{morton:2015} until 100,000 posterior samples had been generated.

We note that for this star, we obtain good agreement between the
light curve derived stellar density and that from our isochrone analysis,
with a ratio of $1.06\pm0.22$. Unaccounted for blend sources would cause this
ratio of these two to deviate from unity \citep{AP} and if the transits were
associated with a completely different star (e.g. in the background) then the
difference could be very large. The fact that this case has a density ratio
within one-sigma of zero (see $\log\Psi$ row in Table~\ref{tab:table_stellar})
thus provides additional support that this is indeed a genuine planet
transiting the target star.

\begin{table*}
\caption{Summary of the stellar parameters calculated from the isochrone analysis for the host star of \toitenb. The parameters below the horizontal line are the physical dimensions of the stars. 
} 
\begin{center}
\begin{tabular*}{\textwidth}{@{\extracolsep{\fill}}lll}
\hline
Parameter & Units & \toiten\ (TOI-1055)\\
\hline
$V$ & $V$-band Magnitude  & $8.68 \pm 0.01$ \\
$J$ & $J$-band Magnitude  & $7.518 \pm 0.023$ \\
$H$ & $H$-band Magnitude  & $7.231 \pm 0.047$ \\
$K$ & $K$-band Magnitude  & $7.150 \pm 0.027$ \\
$T_{\mathrm{eff}}$ & Effective Temperature (K) & $5788\pm44$ \\
Fe/H & Iron-to-Hydrogen Ratio & $-0.023\pm0.050$ \\
$\log(g)$ & Surface Gravity & $4.50\pm0.03$ \\
$\pi$ & Parallax & $17.516\pm0.066$ \\
\hline
$d$ & Distance (pc) & $57.06_{-0.24}^{+0.25}$ \\
$M_{\star}$ & Stellar Mass $(M_{\odot})$ & $1.031_{-0.026}^{+0.025}$ \\
$R_{\star}$ & Stellar Radius $(R_{\odot})$ & $0.985_{-0.026}^{+0.037}$ \\
$\log_{10}(L)$ & Log Luminosity  & $0.012_{-0.032}^{+0.043}$ \\
Age & Age (Gyr) & $2.6_{-1.2}^{+1.4}$ \\
$\rho_{\star}$ & Stellar Density (g\,cm$^{-3}$) & $1.52_{-0.16}^{+0.13}$ \\
\hline
\label{tab:table_stellar}
\end{tabular*}
\end{center}
\end{table*}

\subsection{Transit Analysis}
\label{sub:transits}

We further improve our understanding of the validated planet by including an analysis
of its transit light curve. We downloaded the 2-minute Pre-Data search
Conditioning (PDC) light curve for source from the Mikulski Archive for Space
Telescopes (MAST). At the time of writing, TOI-1055b had been observed in
Sectors 13 and 27, exhibiting 2 and 1 transits respectively.

The light curve was cleaned of time stamps indicating error codes and
outliers using moving median filter. We then detrended the light curve of
long-term trends following the method marginalization approach described in
\citet{toi216}. As in that paper, the scatter between different model
detrendings is propagated into the updated formal uncertainties on our method
marginalized light curve.

The transit light curve was then fit using a 9-parameter \citet{mandel:2002}
forward model coupled to a multimodal nested sampling algorithm, \multi\
\citep{feroz:2009}. Limb darkening was modeled using a quadratic law but
re-parameterized to the $q_1$-$q_2$ formulation of \citet{q1q2}, to enable
efficient exploration of the parameter volume. We parameterize the rest of
the transit model with seven other terms: the time of transit minimum, $\tau$,
the orbital period, $P$, the impact parameter, $b$, the ratio-of-radii, $p$,
the mean stellar density, $\rho_{\star}$, the orbital eccentricity, $e$,
and the argument of periastron, $\omega$. For many of these, we adopt a
simple uniform prior ($q_1 \in [0,1]$, $q_2 \in [0,1]$,
$\tau \in [\hat{\tau}-0.1,\hat{\tau}+0.1]$,
$P \in [\hat{P}-0.1,\hat{P}+0.1]$,
$b \in [0,2]$,
$p \in [0,1]$).
Note that $\hat{P}$ and $\hat{\tau}$ are the \tess\ reported best-fitting
ephemeris parameters.

Eccentricity and mean stellar density are degenerate in a light curve fit
\citep{AP}, and so we use the stellar mean density derived from our isochrone
analysis (see Section~\ref{sub:isochrones}) as an informative prior.
After trying several different parameteric distributions to describe the
isochrone derived stellar density distribution, we found the following provided a
good approximation: $\rho_{\star} \sim \mathcal{W}[1563,11]$\,kg\,m$^{-3}$
(where $\mathcal{W}$ is a Weibull distribution).
For eccentricity and argument of periastron, we use the same
joint prior as described earlier in Section~\ref{sub:radvelfits}.

\begin{figure}
\includegraphics[width=\columnwidth]{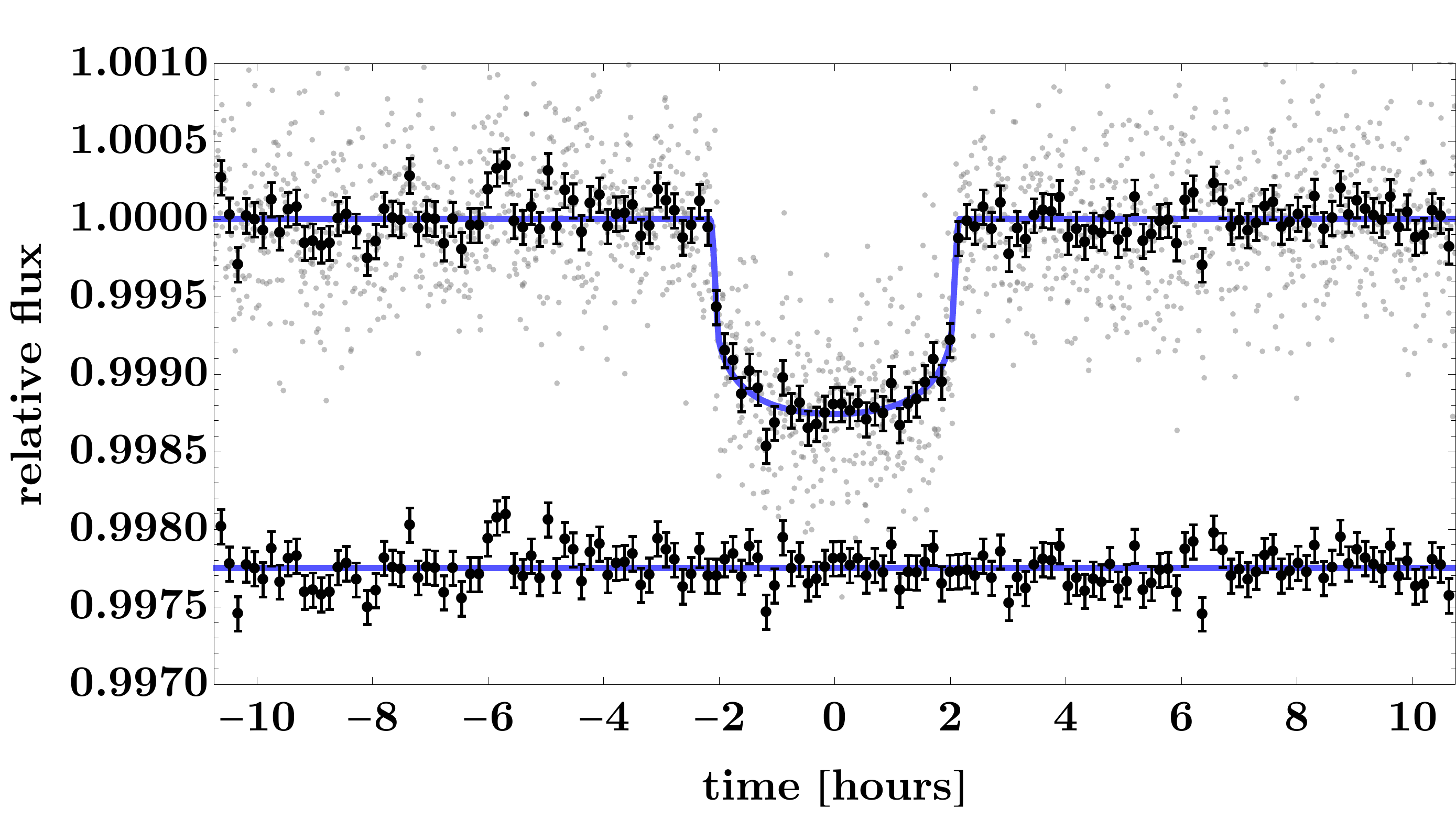}
\caption{
Phase-folded transit light curve of \toitenb\ (TOI-1055b) as
observed by \tess. The black points represent the method marginalized
detrended 2-minute \tess\ photometry and the red line shows the maximum
a-posteriori fit from our regressions. The lower panel shows the residuals
between the two.
}
\label{fig:1055lc}
\end{figure}

We make the posterior samples publicly available for the these this regression
at \href{https://github.com/skypalat/toi_validation}{the aforementioned GitHub
repo}. The maximum \textit{a-posteriori} solution is plotted in
Figure~\ref{fig:1055lc} for \toitenb. The physical parameters implied by this
fit is discussed later in Section~\ref{sub:planetproperties}.

\subsection{Refined \RadVel\ Fits}

Although we have obtained an orbital fit for the radial velocities earlier
in Section~\ref{sub:radvelfits}, that analysis did not include any
eccentricity constraints from the transit fit, since the TOIs remained
unvalidated at that time. Having now validated \toitenb\, we re-run the
\RadVel\ fit for this planet including the eccentricity constraints from the
transit to improve the overall precision in our final system parameters.

This is accomplished by introducing a modification to the \RadVel\ likelihood
function that accounts for this constraint on the orbital eccentricity.
The ratio of light curve derived stellar density (see
Section~\ref{sub:transits}) to that from an independent measure - in our case
from isochrones (see Section~\ref{sub:isochrones}) - directly yields
$\Psi \equiv (1+e \sin\omega)^3 (1-e^2)^{-3/2}$, as shown in
\citet{investigations:2010} (see their Equation~39).

To implement this constraint, in the \logprob\ function of the \RVLikelihood\
class of the module's \likelihoodpy\ file, we added a custom log-likelihood
function describing the agreement between each trial's predicted $\log\Psi$
versus observed $\log\Psi$ (log of the density ratio) value. This
was achieved by using kernel density estimation on the transit $\log\Psi$
posteriors that was then used to tabulate a grid of log-like versus
$\log\Psi$, which was then approximated with a piecewise fourth-order
polynomial with a break at $\log\Psi=0$. We sampled this function in
a test MCMC to ensure it reproduces the eccentricity distribution from
the transits, as expected. This typically helps reduce the amount of time
\RadVel\ spends exploring highly eccentric solutions and keeps the radial
velocity solution in line with that found from the transit analysis
(see Section~\ref{sub:transits}).

Note that since we use the posteriors from the transit fit to construct
the revised \RadVel\ prior, it is not necessary (or indeed allowed) to
include the intrinsic Beta prior and transit bias priors from before,
since these are already baked into the $\log \Psi$ posterior.

\section{Discussion}
\label{sec:discussion}

\subsection{Properties of \toitenb}
\label{sub:planetproperties}

In this work, we report the validation of one planet orbiting \toiten, which
represents a new exoplanet. A summary table of the physical properties is shown
in Table~\ref{tab:table_par}.

\begin{table}
\caption{
Median and $\pm 38.1$\% quantiles of the joint posteriors for
\toitenb's fitted parameters (top) and derived parameters (bottom) $\dagger$:\tess\ BJD is equivalent to BJD - 2457000.
} 
\centering 
\begin{tabular}{c c} 
\hline 
Parameter & Value \\
\hline
$P$ [days] & $17.471278_{-0.000060}^{+0.000058}$ \\
$\tau$ [\tess\ BJD] $\dagger$ & $1661.06315_{-0.00077}^{+0.00078}$ \\
$p \equiv R_P/R_{\star}$ & $0.03300_{-0.00059}^{+0.00063}$ \\
$b$ & $0.32_{-0.20}^{+0.17}$ \\
$\rho_{\star}$ [g\,cm$^{-3}$] & $1.53_{-0.17}^{+0.13}$ \\
$q_{1}$ & $0.38_{-0.17}^{+0.26}$ \\
$q_{2}$ & $0.24_{-0.16}^{+0.30}$ \\
$e$ & $<0.28$ [2\,$\sigma$] \\
$K$ [m\,s$^{-1}$] & $4.9_{-1.0}^{+0.9}$ \\
$\gamma$ [m\,s$^{-1}$] & $5.2_{-1.6}^{+1.5}$; $-3.3_{-1.0}^{+1.1}$ \\
$\dot{\gamma}$ [m\,s$^{-1}$\,yr$^{-1}$] & $-3.2_{-0.7}^{+0.8}$ \\
$\ddot{\gamma}$ [m\,s$^{-1}$\,yr$^{-2}$] & $0.063_{-0.048}^{+0.043}$ \\
$\sigma_{\mathrm{jitter}}$ [m\,s$^{-1}$] & $2.64_{-0.57}^{+0.81}$; $3.61_{-0.55}^{+0.67}$ \\
\hline
$R_{P}$ [$R_{\oplus}$] & $3.55_{-0.12}^{+0.15}$ \\
$M_{P}$ [$M_{\oplus}$] & $19.7_{-3.9}^{+4.0}$  \\
$\rho_{P}$ [g\,cm$^{-3}$] & $2.39_{-0.54}^{+0.57}$ \\
$i$ [$^{\circ}$] & $89.33_{-0.40}^{+0.41}$  \\ 
$a/R_{\star}$ & $29.1_{-1.1}^{+0.8}$ \\
$a$ [AU] & $0.1334_{-0.0061}^{+0.0062}$ \\
$T_{14}$ [hours] & $4.36_{-0.51}^{+0.23}$ \\
$T_{23}$ [hours] & $4.04_{-0.50}^{+0.25}$ \\
$\tilde{T}$ [hours] & $4.20_{-0.50}^{+0.24}$ \\
$u_{1}$ & $0.59_{-0.21}^{+0.17}$ \\
$u_{2}$ & $0.01_{-0.24}^{+0.33}$ \\
$\log\Psi$ & $0.06_{-0.22}^{+0.17}$ \\
$S$ [$S_{\oplus}$] & $58.1_{-3.9}^{+5.3}$\\
$T_{\mathrm{blackbody}}$ [K] & $769_{-13}^{+17}$ \\
TSM & $72_{-13}^{+19}$ \\ [1ex]
\hline 
\end{tabular}
\label{tab:table_par} 
\end{table}

\toitenb\ orbits the G2V host star \toiten\ located $d=(57.37\pm0.19)$\,pc away
in the Telescopium constellation. This star is notably bright in both the
optical and infrared at $V=8.67$ and $K=7.15$, and therefore offers favorable
conditions for follow up observations. 
From our isochrone analysis, we determine that \toiten\ is
$(1.03\pm0.051)$\,$M_{\odot}$ and $(1.022\pm0.071)$\,$R_{\odot}$, which imply a
slightly earlier-type than that reported in TIC-8 ($1.04\pm0.14$\,$M_{\odot}$
and $0.975\pm0.055$\,$R_{\odot}$; \citealt{stassun:2019}).
The mass, radius, and spectral type of \toiten\ are remarkably similar to that
of the Sun. In fact, \toiten\ has been the subject of various analyses of
Sun-like stars including those for chemical abundances
\citet{2018ApJ...865...68B}, infrared excess \citet{DaCosta:2017}, and stellar
age compared to chemical composition \citet{2016AA...590A..32T}. Each of these
studies indicates that \toiten\ exhibits the typical properties of a Solar
twin, including having a spectrum very similar to the Sun.

The RV measurements for this star come from HARPS, with 53 measurements
spanning the dates October 13, 2011 to October 21, 2017. We determine that the
quadrature jitter term is approximately 3\,m/s, close to the median formal
uncertainties for the data set of 1.2\,m/s and indicating that the star is
relatively quiet. The target is flagged as having an ``unambiguous'' rotational
modulation by \citet{canto:2020} with a clear periodicity present in the light
curve at 8.8\,days. Regressing a sinusoid to the Sector 13 light curve, we
obtain an amplitude of 260\,ppm, against which there is residual scatter of
420\,ppm - consistent with the median formal uncertainty of 409\,ppm. In Sector
27, we find almost the same periodicity (8.9\,days) of amplitude 240\,ppm,
against which there is residual scatter of 417\,ppm - consistent with the
median formal uncertainty of 374\,ppm. We thus conclude that the star likely
exhibits rotational modulations due to spots, but this activity is small a
${\sim}250$\,ppm and thus generally consistent with a quiet star.

\begin{figure*}
\begin{center}
\includegraphics[width=\textwidth]{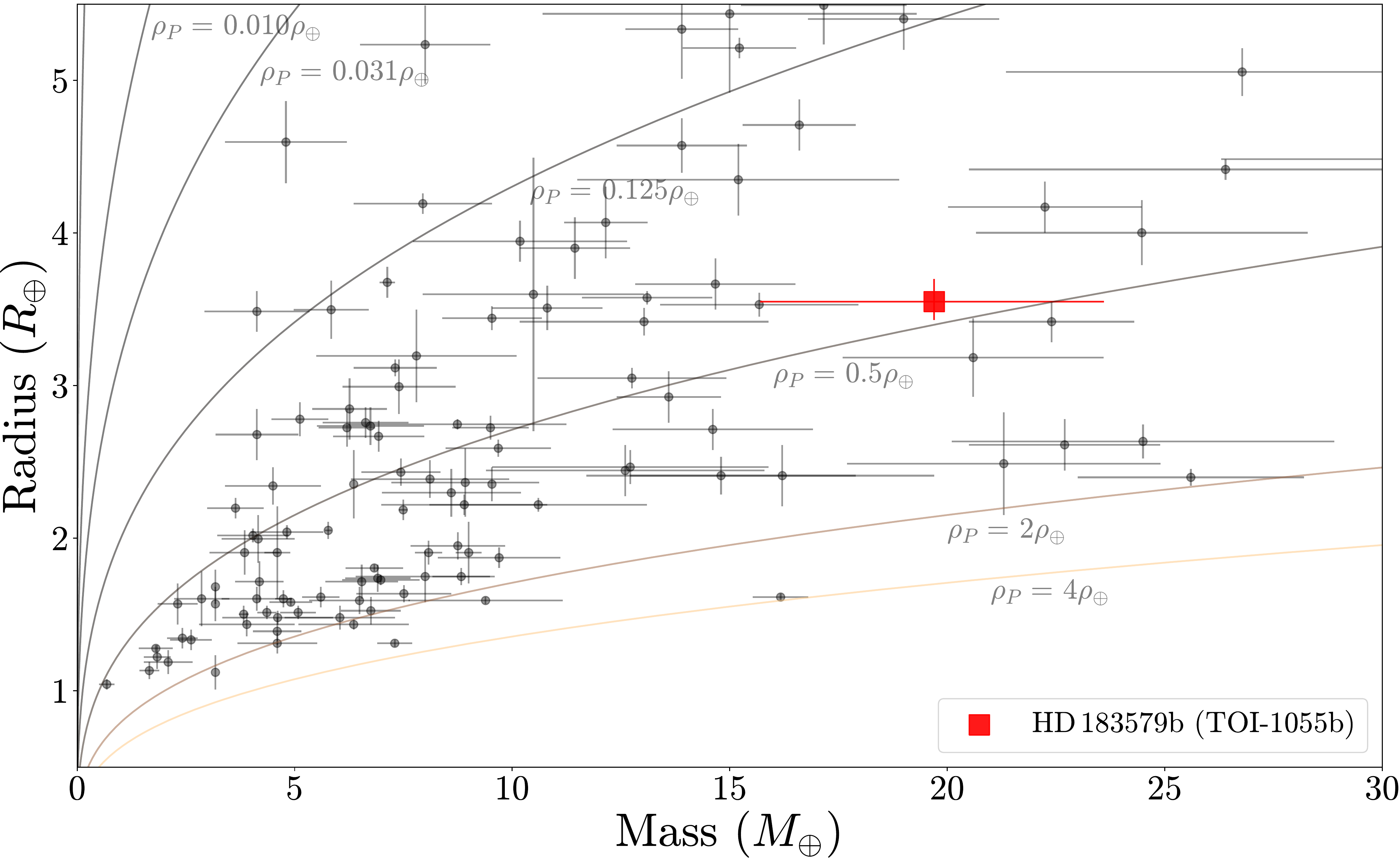}
\caption{
A mass-radius diagram demonstrating where the newly validated planet lies
amongst the population of known planets. \toitenb\ is represented by a red square.
Contour lines indicate levels of constant density. \toitenb\ has dimensions
consistent with a Neptunian exoplanet.
}
\end{center}
\label{fig:mrdiag}
\end{figure*}

For \toitenb, we report a radius of $(3.55\pm0.13)$\,$R_{\oplus}$, thus
placing it firmly in the Neptunian category of \citet{chen:2017}.
We determine a mass for \toitenb\ of $M_P = 19.7_{-3.9}^{+4.0}$\,$M_{\oplus}$,
indicating to a bulk density of $\rho_P = 2.39_{-0.54}^{+0.57}$\,g\,cm$^{-3}$. The planetary mass and radius indicate that \toitenb\ resembles Neptune/Uranus in
bulk density, and perhaps has thus migrated inwards from beyond the ice-line.
Figure \ref{fig:mrdiag} is a standard mass-radius diagram demonstrating where
\toitenb\ falls in a distribution of known planets.

\toitenb\ orbits its host star once every $17.5$\,days at a semi-major axis of
$(0.1334\pm0.0062)$\,AU. From 
the transit morphology, we find that the orbital eccentricity is consistent
with a circular path with a median of $0.14_{-0.10}^{+0.26}$. The FAP of this
being eccentric using the \citet{lucy:1971} test is 37\%, thus favoring a
circular orbit. Further, using the Savage-Dickey ratio \citep{dickey:1971}, we
compute the Bayes factor between an eccentric-to-circular orbit model to
$0.39$ - again indicating a preference for the circular solution. Using just
the transits, we conclude $e<0.66$ to 95.45\% confidence.

The transit posteriors imply a constraint on $\log \Psi = 0.06_{-0.22}^{+0.17}$
(median and standard deviation) which is, recall, propagated as a prior
constraint on eccentricity in our \RadVel\ fits. From \RadVel, the eccentricity
constraints are slightly improved by the inclusion of the RV information,
yielding $e=0.14_{-0.08}^{+0.07}$ - which may suggest some small amount of
eccentricity that offer clues to this planet's past. However,
we caution that neither the Savage-Dickey ratio nor the Lucy-Sweeney test
formally favor an elliptical orbit at this point. We conclude that $e<0.27$
to 95.45\% confidence, and once again remark that an RV trend appears to
indicate an outer body with a significance of $\Delta\mathrm{BIC}>30$.

From our measured mass and radius, we calculate a transmission spectroscopy
metric (TSM) for \toitenb\ using Equations (1) \& (2) of
\citet{2018PASP..130k4401K} to indicate the expected signal-to-noise ratio for
future James Webb Space Telescope (JWST) measurements. Our calculated
TSM$= 72_{-13}^{+19}$ indicates that \toitenb\ is a promising object for
future JWST observations.

\subsection{The Use of Archival RVs}

Using exclusively publicly available resources, we were able to validate and
characterize the physical properties of one Neptune sized exoplanet, \toitenb.
Thus, existing data advances \tess\ primary objective of measuring the masses
and radii of 50 small ($<4$\,$R_{\oplus}$) exoplanets \citet{ricker:2015}.

The planet itself is a fascinating world that will likely be amongst the rare
planets observed by JWST, thanks to its small size and excellent observability.
But, we would also like to highlight that
the technique used to validate this object could be extended and utilized in
the future. For example, in this work we did not consider multiple planet
systems as the FAP scoring system would require some modification to handle the
multiple periodicities. Nevertheless, multiples are intrinsically more likely
to be genuine planets \citep{lissauer:2012} and thus would need less of a nudge
in a probability-sense to become validated planets. Our work highlights the
great power of legacy RV surveys in synergy with active missions, such as
\tess. And, it demonstrates that the knee-jerk reaction of going to the
telescope to get new data is not always necessary, in some cases existing
archives may in fact already serve the desired goal.

\section*{Acknowledgements}

DK acknowledges support from Columbia's Data Science Institute. The Cool Worlds group thanks to Tom Widdowson, Mark Sloan, Douglas Daughaday, Andrew Jones, Jason Allen, Marc Lijoi, Elena West, Tristan Zajonc, Chuck Wolfred, Lasse Skov, Geoff Suter, Max Wallstab, Methven Forbes, Stephen Lee, Zachary Danielson \& Vasilen Alexandrov.

This research has made use of the NASA Exoplanet Archive, which is operated by the California Institute of Technology, under contract with the National Aeronautics and Space Administration under the Exoplanet Exploration Program.

This research has made use of the SIMBAD database, operated at CDS, Strasbourg, France.

This work has made use of data from the European Space Agency (ESA) mission \href{https://www.cosmos.esa.int/gaia}{\textit{Gaia}}, processed by the Gaia Data Processing and Analysis Consortium (\href{https://www.cosmos.esa.int/web/gaia/dpac/consortium}{DPAC}). Funding for the DPAC has been provided by national institutions, in particular the institutions participating in the Gaia Multilateral Agreement.

This paper includes data collected with the \tess\ mission, obtained from the MAST data archive at the Space Telescope Science Institute (STScI). Funding for the \tess\ mission is provided by the NASA Explorer Program. STScI is operated by the Association of Universities for Research in Astronomy, Inc., under NASA contract NAS 526555.

We acknowledge the use of public TESS Alert data from pipelines at the TESS Science Office and at the TESS Science Processing Operations Center.

We are deeply grateful to the HARPS team at Observatoire de Genève, Observatoire de Haute-Provence, Laboratoire d’Astrophysique de Marseille, Service d’Aéronomie du CNRS, Physikalisches Institut de Universität Bern, ESO La Silla, and ESO Garching, who built and maintained the HARPS instrument, and were generous enough to make the data public. 

Research at the Lick Observatory is partially supported by a generous gift from Google. Some of the data presented herein were obtained at the W. M. Keck Observatory, which is operated as a scientific partnership among the California Institute of Technology, the University of California, and NASA. The Observatory was made possible by the generous financial support of the W.M. Keck Foundation.

We thank all of the observers who spent countless nights using both the HARPS and LCES facilities to collect the data presented here and all of the PIs who submitted telescope proposals year after year to allow the acquisition of these data.

Finally, the authors wish to recognize and acknowledge the very significant cultural role and reverence that the summit of Maunakea has always had within the indigenous Hawaiian community. We are most fortunate to have benefited from the observations obtained from this mountain.

\textit{Facilities}: Keck:I (HIRES), \tess, HARPS

\textit{Software}: \emcee\ \citep{dfm:2013}, \multi\ \citep{feroz:2009}, \RadVel\ \citep{fulton:2018}, \forecaster\ \citep{chen:2017}, \isochrones\ \citep{morton:2015}






\end{document}